\documentclass[proof]{pasj01}
\Received{$\langle$reception date$\rangle$}
\Accepted{$\langle$acception date$\rangle$}
\Published{$\langle$publication date$\rangle$}

\begin{document}

\title{{LAMOST} J~2217+2104: a new member of carbon-enhanced extremely metal-poor
stars with excesses of Mg and Si}
\author{Wako \textsc{Aoki}\altaffilmark{1}\altaffilmark{2}}
\altaffiltext{1}{National Astronomical Observatory, 
 2-21-1 Osawa, Mitaka, Tokyo 181-8588, Japan }
\email{aoki.wako@nao.ac.jp}

\author{Tadafumi \textsc{Matsuno}\altaffilmark{2}}
\altaffiltext{2}{Department of Astronomical Science, School of Physical Sciences, The Graduate University of Advanced Studies (SOKENDAI), 2-21-1 Osawa, Mitaka,
Tokyo 181-8588, Japan}

\author{Satoshi \textsc{Honda}\altaffilmark{3}}
\altaffiltext{3}{Center for Astronomy, University of Hyogo, 407-2, Nishigaichi, Sayo-cho, Sayo, Hyogo 679-5313, Japan}

\author{Miho \textsc{Ishigaki}\altaffilmark{4}}
\altaffiltext{4}{Kavli Institute for the Physics and Mathematics of the Universe (WPI), The University of Tokyo, Kashiwa, Chiba 277-8583, Japan}

\author{Haining \textsc{Li}\altaffilmark{5}}
\altaffiltext{5}{Key Lab of Optical Astronomy, National Astronomical
  Observatories, Chinese Academy of Sciences, A20 Datun Road,
  Chaoyang, Beijing 100012, China}

\author{Takuma \textsc{Suda}\altaffilmark{6}}
\altaffiltext{6}{Research Center for the Early Universe, Graduate School of Science, University of Tokyo, Hongo, Tokyo 113-0033, Japan}

\author{Yerra Bharat \textsc{Kummar}\altaffilmark{5}}




\maketitle

\begin{abstract}

We report on the elemental abundances of the carbon-enhanced
metal-poor (CEMP) star J~2217+2104 discovered by our metal-poor star
survey with LAMOST and Subaru.  This object is a red giant having
extremely low Fe abundance ([Fe/H]=-4.0) and very large enhancement of
C, N, and O with excesses of Na, Mg, Al, and Si. This star is a new
example of a small group of such CEMP stars identified by previous
studies. We find a very similar abundance pattern for O-Zn in this
class of objects that shows enhancement of elements up to Si and
normal abundance of Ca and Fe-group elements. Whereas the C/N ratio is
different among these stars, the (C+N)/O ratio is similar. This
suggests that C was also yielded with similar abundance ratios
relative to O--Zn in progenitors, and was later affected by the CN-cycle. 
By contrast, the heavy neutron-capture elements Sr and Ba are
deficient in J~2217+2104, compared to the four objects in this class
previously studied. This indicates that the neutron-capture process in
the early Galaxy, presumably the r-process, has no direct connection
to the phenomenon that has formed such CEMP stars. Comparisons of the
abundance pattern well determined for such CEMP stars with those of
supernova nucleosynthesis models constrain the progenitor mass to be
about 25~M$_{\odot}$, which is not particularly different from typical
mass of progenitors expected for extremely metal-poor stars in
general.

\end{abstract}

\section{Introduction}

Extremely Metal-Poor (EMP) stars ([Fe/H]$\lesssim -3$)\footnote{[A/B]
  = $\log(N_{\rm A}/N_{\rm B}) -\log(N_{\rm A}/N_{\rm B})_{\odot}$,
  and $\log\epsilon_{\rm A} =\log(N_{\rm A}/N_{\rm H})+12$ for
  elements A and B.} in the Milky Way are regarded as old low-mass
stars formed in the very early stage of the Galaxy evolution recording
the yields of first generations of massive stars \citep{frebel15}. A
remarkable feature found in EMP stars is the high frequency of
Carbon-Enhanced Metal-Poor (CEMP) stars ([C/Fe]$>+0.7$\footnote{We
  adopt [C/Fe]$>+0.7$ as the definition of CEMP stars in this paper
  following the definition of SAGA database \citep{suda08}, whereas
  [C/Fe]$>+1.0$ is adopted in some other papers.}). A large fraction
of CEMP stars with [Fe/H]$>-3$ also show excesses of heavy
neutron-capture elements like Ba, a signature of contribution by the
s-process nucleosynthesis in AGB stars through mass transfer across
binary systems (so-called CEMP-s stars).  By contrast, CEMP stars with
no excess of Ba (CEMP-no stars) are dominant at [Fe/H]$<-3$ (e.g.,
\cite{aoki07}). CEMP stars with such low metallicity could record
the nucleosynthesis yields of a kind of supernova that is particularly
effective in the early Galaxy. The high frequency of such objects
could be also related to the cooling process of star forming
clouds. The excess of C, as well as O that is also overabundant in
some CEMP stars, could result in efficient gas cooling, which is
required for low-mass star formation, even at very low metallicity
\citep{bromm03, norris13}.

Most of the CEMP-no stars do not show any anomaly in abundance ratios
except for the light elements C, N, and O (e.g., \cite{aoki07,
  ito09}). The [C/Fe] values of the CEMP-no stars show a wide distribution,
including many objects that have a ratio close to the criterion of
CEMP stars ([C/Fe]$\sim +1$; Yoon et al. 2016). These observational
features could suggest a connection of some fraction of CEMP-no stars with C-normal stars.

There are, however, a small number of CEMP-no stars that show large
excesses of some $\alpha$-elements, which well separate them from
C-normal EMP stars. The first example of such a star, CS~22949--037,
was reported by \citet{mcwilliam95}, and was studied in more detail by
\citet{norris01}, \citet{norris02} and \citet{depagne02}. This is a
CEMP star with [Fe/H]$\sim -4$ showing no significant excess of
neutron-capture elements, but it exhibits clear excesses of Mg and Si,
as well as odd elements with similar atomic mass (Na, Al). Another EMP
star having a similar abundance pattern, CS~29498--043 ([Fe/H]$\sim
-3.5$), was reported by \citet{aoki02}, who suggested the existence of
a class of CEMP-no stars with significant excesses of Mg and Si in the
lowest metallicity range. Two more CEMP-no stars having similarly
large excess of Mg and Si have been reported by
\citet{cohen13} and \citet{yong13}: HE~1012--1510 ([Fe/H]$=-3.5$) and
HE~2139--5432 ([Fe/H]$=-4.0$). More recently, \citet{bonifacio18}
report another EMP star, SDSS~J~1349+1407, having [Fe/H]$=-3.6$ and
large excesses of C, Na, and Mg, although the number of elements for
which abundances are measured is still small for this faint, unevolved object.  These stars are well separated from other stars
with similar Fe abundances in the $\alpha$/Fe abundance ratios (e.g.,
[Mg/Fe]; \cite{norris13}). Given the fact that $\alpha$-elements are
efficiently produced during the evolution of massive stars and their
supernova explosions, these CEMP stars could be regarded as a record
of nucleosynthesis by the first generation of massive stars.

Large excesses of C, N, O and Mg are also found in the ``Hyper
Metal-Poor'' ([Fe/H]$<-5$) star HE~1327--2326
\citep{frebel05}. Although the over-abundances of C, N, and O are much
more significant in this object than those of the above CEMP stars,
the overall abundance trend could suggest that these stars
  have similar origins.

The distributions of abundance ratios for CEMP-no stars were discussed
in detail by \citet{norris13}, in which stars with excesses of Na, Mg
and Si are focused. As possible origins of these objects, so-called
faint supernovae explained by the ``mixing fall-back'' models
\citep{umeda03} and mass-loss from rotating massive stars
\citep{meynet06} are argued. To discriminate these possibilities,
detailed abundance pattern of $\alpha$-elements up to Ti is a key. The
sample of such stars is, however, still very limited as mentioned
above, even though EMP stars have been intensively searched for in the
past two decades. This could be due to the fact that such EMP stars
([Fe/H]$\lesssim -3.5$) are generally very rare. Taking account of the
importance of such objects in the understanding of formation and
evolution of the first generations of stars, further searches for new
examples and detailed abundance studies are desired.

Here we report on the discovery of an Ultra Metal-Poor
([Fe/H]$\lesssim -4$) star, {LAMOST} J~221750.59+210437.2
(J~2217+2104), that shows large excesses of $\alpha$-elements as well
as C, and discuss the detailed abundance pattern of this class of
objects.

\section{Observations and measurements}

J~2217+2104 was discovered by the medium resolution ($R\sim 1800$)
spectroscopy with the Large Sky Area Multi-Object Fiber
Spectroscopic Telescope (LAMOST) in its regular survey \citep{cui12, zhao12}
as a candidate EMP star. The stellar parameters
estimated from the LAMOST spectrum indicate that this object is an
EMP  giant with [Fe/H]$<-3.5$. Strong 
CH molecular bands are identified in the LAMOST spectrum.

High-resolution spectra of this object were obtained with the Subaru
Telescope High Dispersion Spectrograph (HDS; \cite{noguchi02}) in
observing programs for follow-up spectroscopy of metal-poor star
candidates found with LAMOST (\cite{li15}). The first spectrum was
obtained in November 2015 with a short exposure (so-called
``snapshot'' spectroscopy). High-resolution spectra with longer
exposures were obtained in August 2017 with a resolving power of
$R=60,000$ by two wavelength setups, covering 3500-5200~{\AA} and
4030-6800~{\AA}, respectively. Details of the observations are given
in table~\ref{tab:obs}.

Standard data reduction procedures were carried out with the IRAF
echelle package\footnote{IRAF is distributed by the National Optical
  Astronomy Observatories, which is operated by the Association of
  Universities for Research in Astronomy, Inc. under cooperative
  agreement with the National Science Foundation.}. The wavelength
shift due to Earth's orbital motion is corrected using the IRAF task {\it
  rvcor}.

We measure equivalent widths by fitting a Gaussian profile. Line
data for spectral features are taken from previous studies on very
metal-poor stars (e.g., \cite{aoki13}). The measured equivalent widths
are given in table~2, together with the line data used in the abundance
analysis.

The Fe {\small I} lines for which equivalent widths are measured are
also used to measure radial velocities. The heliocentric radial
velocities derived from these Fe {\small I} lines are given in
table~\ref{tab:obs}. The random error in the measurement is estimated
to be $\sigma_{v} N^{-1/2}$, where $\sigma_{v}$ is the standard
deviation of the derived values from individual lines, and $N$ is the
number of lines used. The random errors are
0.03--0.04~km~s$^{-1}$. The errors due to the instability of the
instrument corresponding to, e.g., temperature variations, are usually
larger ($\sim 0.5$~km~s$^{-1}$) than the above errors.  The radial
velocity obtained from the LAMOST spectrum and the value provided by
Gaia DR2 are also given in the table. Given the errors of radial
velocity measurements, no signature of radial velocity variation is
found in the data currently available.

Photometry data ($B=14.697$, $V=13.388$, $J=11.304$ and $K=10.644$) are
taken from APASS \citep{henden16} and 2MASS \citep{cutri03}. The $V$
magnitude of this object indicates that this is the brightest CEMP
star with large excess of Mg and Si known to date.

The parallax provided by the Gaia DR2 ($0.0004\pm 0.0259$ mas) is
still uncertain and is not useful to constrain the distance and
surface gravity.

\section{Abundance analysis and results}\label{sec:cc}

We determine the elemental abundances of J~2217+2104 by 1D/LTE standard
analysis and spectrum synthesis techniques using model atmospheres of
the ATLAS NEWODF grid (\cite{castelli97}). To determine the stellar
parameters, photometry is also used.

\subsection{Stellar parameters and abundance analysis}

We estimate the effective temperature ($T_{\rm eff}$) from the colors
using the temperature scale of \citet{alonso99} for giant stars. We
assume [Fe/H]$=-3$ in the calculation of $T_{\rm eff}$, following
\citet{ryan99}. The reddening ($E(B-V)=0.054$) is adopted from
\citet{schlafly11}. The derived $T_{\rm eff}$ from $(V-K)_{0}$ is
4494~K, which agrees with those from $(B-V)_{0}$ (4567~K) and from
$(J-K)_{0}$ (4615~K) within the uncertainties of the estimates. We
adopt $T_{\rm eff}=4500$~K for the abundance analysis, taking account
of the sensitivity of $V-K$ to $T_{\rm eff}$ and uncertainty due to
errors of photomety data and the temperature scale.

Following the usual abundance analysis procedure, the surface gravity
($g$) and micro-turbulent velocity ($v_{\rm turb}$) are determined as
the Fe abundances derived from Fe {\small I} and Fe {\small II} lines
are consistent, and as there remains no trend of Fe abundances from
individual Fe {\small I} lines as a function of the line strengths,
respectively. For the analysis, 53 Fe {\small I} lines and 4 Fe
{\small II} lines are used. Errors of the Fe abundance measurements of
0.1~dex results in an uncertainty in $\log g$ of 0.2~dex. 
The derived $\log g$ and $v_{\rm turb}$
are 0.9~dex and 2.3~km~s$^{-1}$, respectively.

The derived Fe abundances from individual Fe {\small I} lines show
a correlation with the lower excitation potential
($\delta$[Fe/H]/$\delta \xi = -0.14$~dex/eV). The slope is shallower
if the lines with the lowest excitation potential ($<1$~eV) are
excluded. Such a trend is, however, usually found in the analysis of
extremely metal-poor giants adopting $T_{\rm eff}$ estimated from
colors (e.g., \cite{frebel13}). Hence, we adopt the $T_{\rm eff}$
estimated from the colors with no correction.

We estimate the uncertainties of $T_{\rm eff}$, $\log g$, [Fe/H] and
$v_{\rm turb}$ to be 100~K, 0.3~dex, 0.3~dex, and 0.3~km~s$^{-1}$,
respectively, from the photometry errors and scatter of the Fe
abundances derived from individual Fe lines.

\subsection{Abundance analysis}

The abundances of elements other than C and N listed in table 3 are
determined by equivalent widths analysis. For the analysis of the
  Sc II, Mn I, Co I and Ba II lines, the effect of hyperfine
splitting is included, using the wavelengths and the fraction of
  transition probability of each splitted line in the Kurucz's line
  data\footnote{http://kurucz.harvard.edu/linelists.html}. The
isotope ratios of the solar-system r-process component are assumed for
Ba. The effect is 0.11~dex in the Mn abundance that is determined
  by a single line with moderate strength, whereas the effects are
  minor (0.02 -- 0.03~dex) in the abundances of the other three
  elements.

\begin{figure} 
\begin{center}
  \includegraphics[width=8cm]{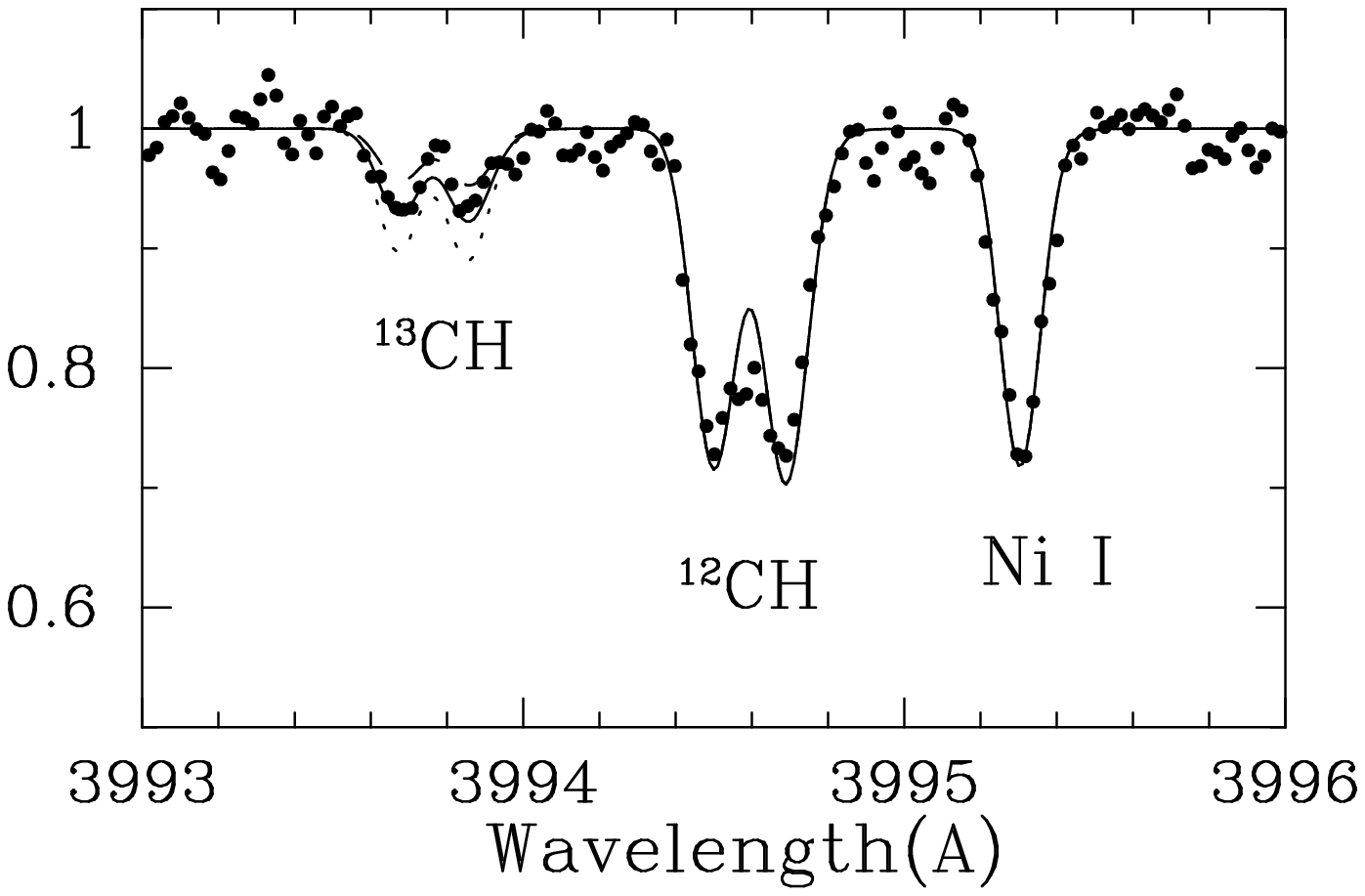} \\
  \includegraphics[width=8.5cm]{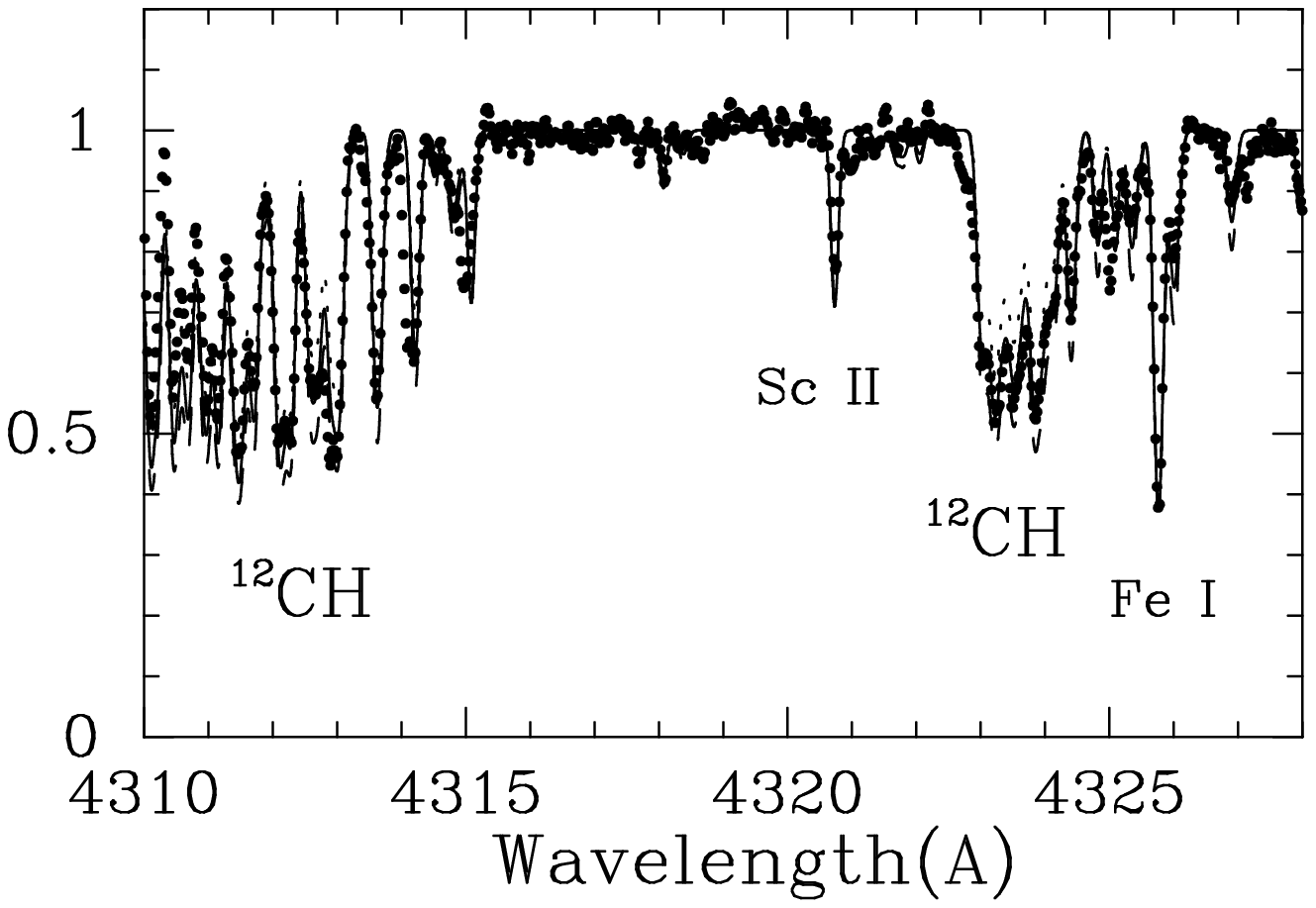} \\
 \end{center}
 \caption{Spectral features of CH molecule in J~2217+2104. (upper) The CH absorption bands with synthetic spectra for [C/Fe]=1.03$\pm 0.2$. (lower) The $^{12}$CH and  $^{13}$CH molecular lines around 3994~{\AA} with synthetic spectra for $^{12}$C/$^{13}$C=4,6, and 10.}
\label{fig:sp}
\end{figure}

The C and N abundances are determined by spectrum synthesis
technique for the CH 4323~{\AA} and CN 3883~{\AA} bands, using the line
lists of \citet{masseron14} for CH and \citet{sneden14} for CN.
The wavelength range of the CH band is shown in figure~\ref{fig:sp}.

The C isotope ratio ($^{12}$C/$^{13}$C) is estimated to be 6 from the
CH molecular lines around 4000~{\AA}. An example of the spectral
feature is depiced in figure~\ref{fig:sp}.

The O abundance is determined from the [O {\small I}] absorption line
at 6300~{\AA}. The stellar absorption line does not overlap with the
telluric features ([O {\small I}] emission line and O$_{2}$ absorption
lines; see \cite{aoki04}) in the spectrum obtained on August 5,
2017. We measure the [O {\small I}] equivalent width with no
correction of telluric features.

No Zn {\small I} line is detected in the spectrum. The upper limit of
Zn abundance ([Zn/Fe]$<0.8$) is estimated from the Zn {\small I}
4722~{\AA} line adopting an upper limit of the equivalent width of
3~m{\AA} at 2$\sigma$ level. We note that the Zn {\small I} 4811~{\AA}
line is not available in the spectrum due to CCD bad columns.

The results of the abundance measurements are given in table 3. The
solar abundances of \citet{asplund09} are adopted to calculate the
[X/Fe] values. The error of the abundance of each element is also
given in the table. The error is obtained by adding in quadrature the
random error and errors due to the uncertainties of stellar
parameters.  The random errors in the measurements are estimated to be
$\sigma N^{-1/2}$, where $\sigma$ is the standard deviation of derived
abundances from individual lines, and $N$ is the number of lines
used. The $\sigma$ of Fe {\small I} ($\sigma_{\rm Fe I}$) is adopted
in the estimates for element X for which the number of lines
  available in the analysis ($N_{\rm X}$) is small (i.e. the error is
$\sigma_{\rm Fe I} N_{\rm X}^{-1/2}$). The errors due to the uncertainty of
the atmospheric parameters are estimated for a giant, for $\delta
T_{\rm eff}= 100$~K, $\delta \log g=0.3$, and $\delta v_{\rm
  turb}=0.3$~km~s$^{-1}$. The errors due to uncertainties of stellar
parameters are similar to the previous estimates for extremely
metal-poor giants (e.g., \cite{aoki04}).

\section{Discussion}

\subsection{Overall abundance pattern of CEMP stars with excesses of Mg and Si}

\begin{figure}
 \begin{center}
  \includegraphics[width=15cm]{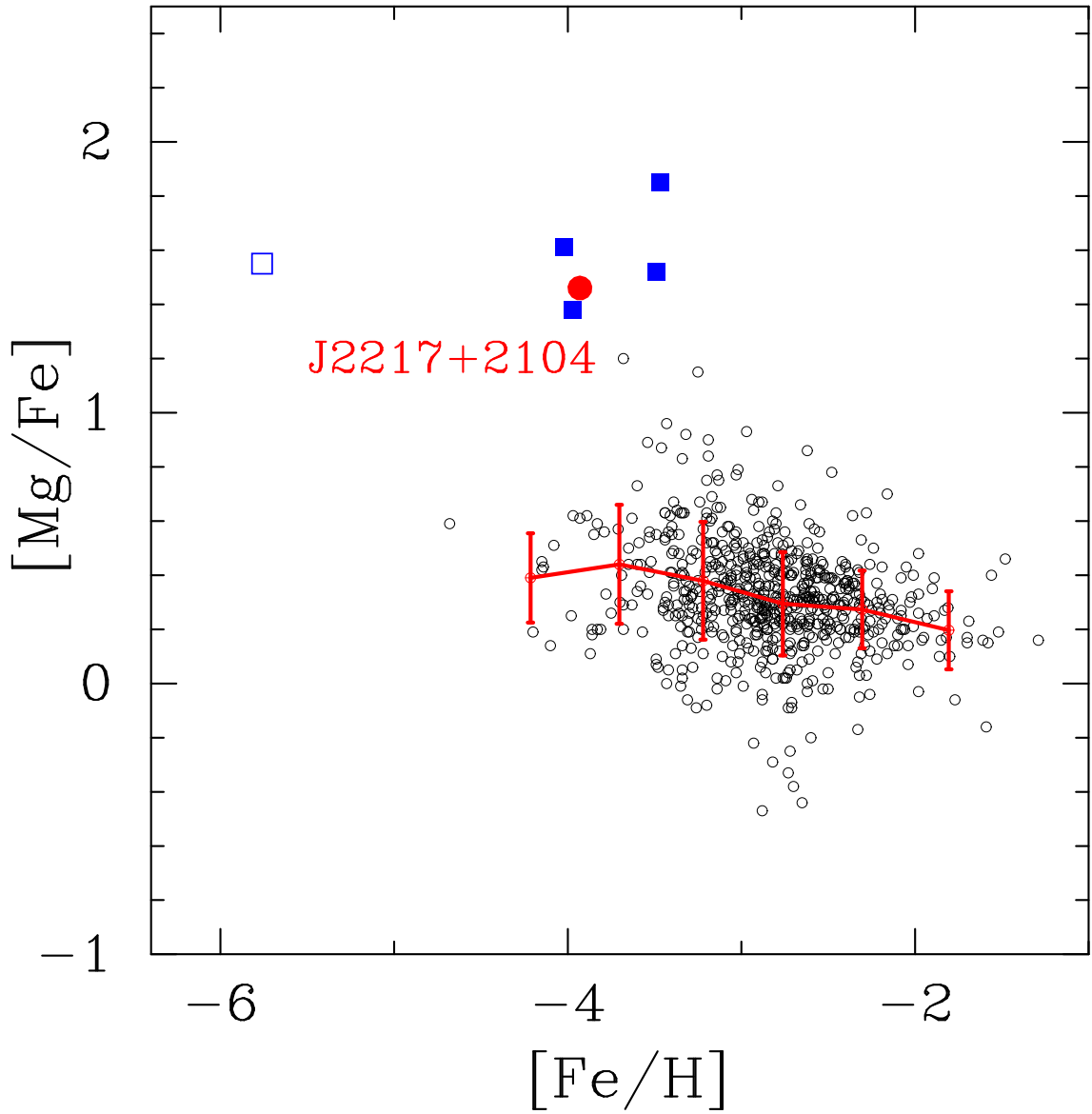}
 \end{center}
 \caption{[Mg/Fe] as a function of [Fe/H] for J~2217+2104 (the red
   filled circle) and other stars.  Abundance data of the four CEMP
   stars with excesses of Mg and Si (blue filled squares) and
   HE~1327--2326 (open square) are adopted from
   \citet{norris13}. Abundance data of other stars (small circles) are
   taken from literature (see text). CEMP stars with excesses of
   neutron-capture elements, mostly CEMP-s stars, are excluded. The
   solid line and bars indicate averages and standard deviations of
   [Mg/Fe] values of 0.25~dex bins of [Fe/H]. In the calculations of
   the averages and standard deviations, the five CEMP stars with
   excesses of $\alpha$-elements are excluded.}\label{fig:mgfe}
\end{figure}

Figure~\ref{fig:mgfe} shows the abundance ratios of [Mg/Fe] as a
function of [Fe/H]. Abundance data of J~2217+2104 and the other four
CEMP stars with Mg excess CS~22949-037, CS~29498-043, HE~1012-1540 and
HE~2139-5432 are shown by the red filled circles and blue filled
squares, respectively. The abundance data for the four objects are
adopted from \citet{norris13}, who compiled the results obtained by
the analysis of \citet{yong13}. The abundance data of the other stars
(small circles) are taken from literature
\citep{cayrel04,honda04,barklem05,lai08,yong13,cohen13,aoki13,jacobson15,hansen15}
through the SAGA database \citep{suda17}. The solid line connects the
averages of [Mg/Fe] (and [Fe/H]) of the objects in 0.25~dex bins of
[Fe/H], and the bar indicates the standard deviations of the [Mg/Fe]
values. In the calculations of the averages and standard deviations,
the five CEMP stars with Mg excess are excluded.

Although statistically significant scatter exists in the [Mg/Fe] ratios of
EMP stars, the large excess of [Mg/Fe] in the five
objects with [Fe/H]$\sim -4$ is distinct. This
demonstrates that the five stars are well separated from the bulk of
the EMP stars that have [Mg/Fe]$\sim +0.4$ on average.

\begin{figure}
 \begin{center}
  \includegraphics[width=15cm]{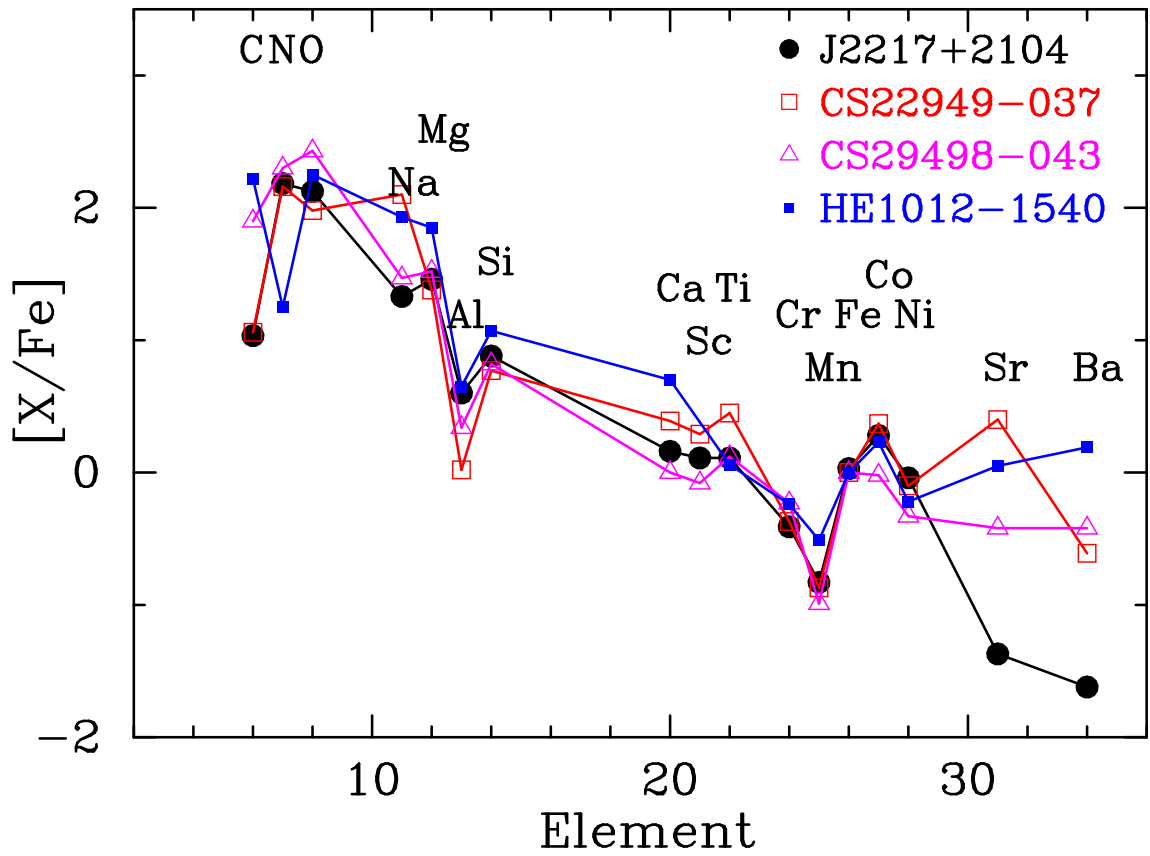}
 \end{center}
 \caption{Abundance patterns of CEMP stars with large excesses of Mg and Si. The object names and corresponding symbols are presented in the panel.}\label{fig:pattern}
\end{figure}

Figure \ref{fig:pattern} shows the overall abundance pattern of
J~2217+2104, and those of other three CEMP stars previously
reported. HE~2139-5432 is not included in the plot, because the number
of elements for which abundance measurements were reported is
relatively small.  The figure demonstrates that the abundance patterns
of the elements between O and Ni of these four stars are very similar,
considering the measurement errors of typically 0.2~dex.
The similarity of the abundance patterns suggests
that the progenitors of these objects, which are expected to be a
specific type of supernova explosions of massive stars, are also
similar. Before discussing the possible origin of the abundance
pattern of these stars, we inspect the abundances of C and N, as well
as the neutron-capture elements Sr and Ba.

\subsection{C and N abundances}\label{sec:cn}

The five objects that show large excesses of Mg and Si identified in
the previous subsection also have very high abundance ratios of C/Fe
and N/Fe. The C/N abundance ratios are, however, different between
stars. The abundance ratios of related elements are given in
table~\ref{tab:ratios} for the five objects. The data of HE~1327--2326
are also given for comparison purposes. The over-abundance of N is
larger than that of C in J~2217+5104, CS~22949--037, and CS~29498--043,
whereas the over-abundance of C is larger in the other two stars.

The C and N abundances could be affected by the CN-cycle, but the
total abundance of the two elements should be preserved during the
process. The total abundance of C and N of all of the five stars are
very high ([(C+N)/Fe]$>+1.6$) although the carbon abundances of
J~2217+2104 and CS~22949--037 ([C/Fe]$\sim +1$) are close to the
criterion of CEMP stars.

The ratio of the total abundance of C and N with respect to
the O abundance (i.e., [(C+N)/O]) is also given in the table.
Interestingly, the (C+N)/O ratios in the five objects are very
similar. The average and the standard deviations of the ratios are
$<$[(C+N)/O]$>=-0.42$ and $\sigma$([(C+N)/O])$=0.19$,
respectively. This suggests that C has been enhanced similarly to O in
these objects, and a portion of the enhanced C is transformed into N
by the CN-cycle, resulting in a variation between the stars.

The C and N abundances could be affected by the CN-cycle in the
progenitors of these objects, which would be massive stars that
produced large excesses of Mg and Si. The CN-cycle is, however, also
effective in low-mass red giants that we are currently
observing. Indeed, a large fraction of highly evolved red giants with
very low metallicity show very low C and high N abundances (``mixed
stars''; \cite{spite05}). The two objects having relatively high C/N
ratios among the five objects are warmer ($T_{\rm eff} > 5400$~K) than
the other three objects (table~\ref{tab:ratios}). This suggests that
the variation of the C/N ratios could be due to the CN-cycle during
the low-mass star evolution. The low $^{12}$C/$^{13}$C ratios in the
three cool stars and non-detection of $^{13}$C in the others 
(table~\ref{tab:ratios}) support this interpretation.

The [C/O] ratios in metal-poor Damped Lyman-$\alpha$ systems are
similar to the [(C+N)/O] values of CEMP stars studied
here. \citet{cooke17} argue that the source of the C and O in the
systems are supernova explosions of $\sim 20$M$_{\odot}$ based on
nucleosynthesis models for first stars and supernova explosions
in which the C/O ratio is sensitive to the mass of progenitors
\citep{heger10}. The constraint on progenitor mass obtained by
elemental abundances are discussed in \S 4.4.

\subsection{Neutron-capture elements}

\begin{figure} 
\begin{center}
  \includegraphics[width=8cm]{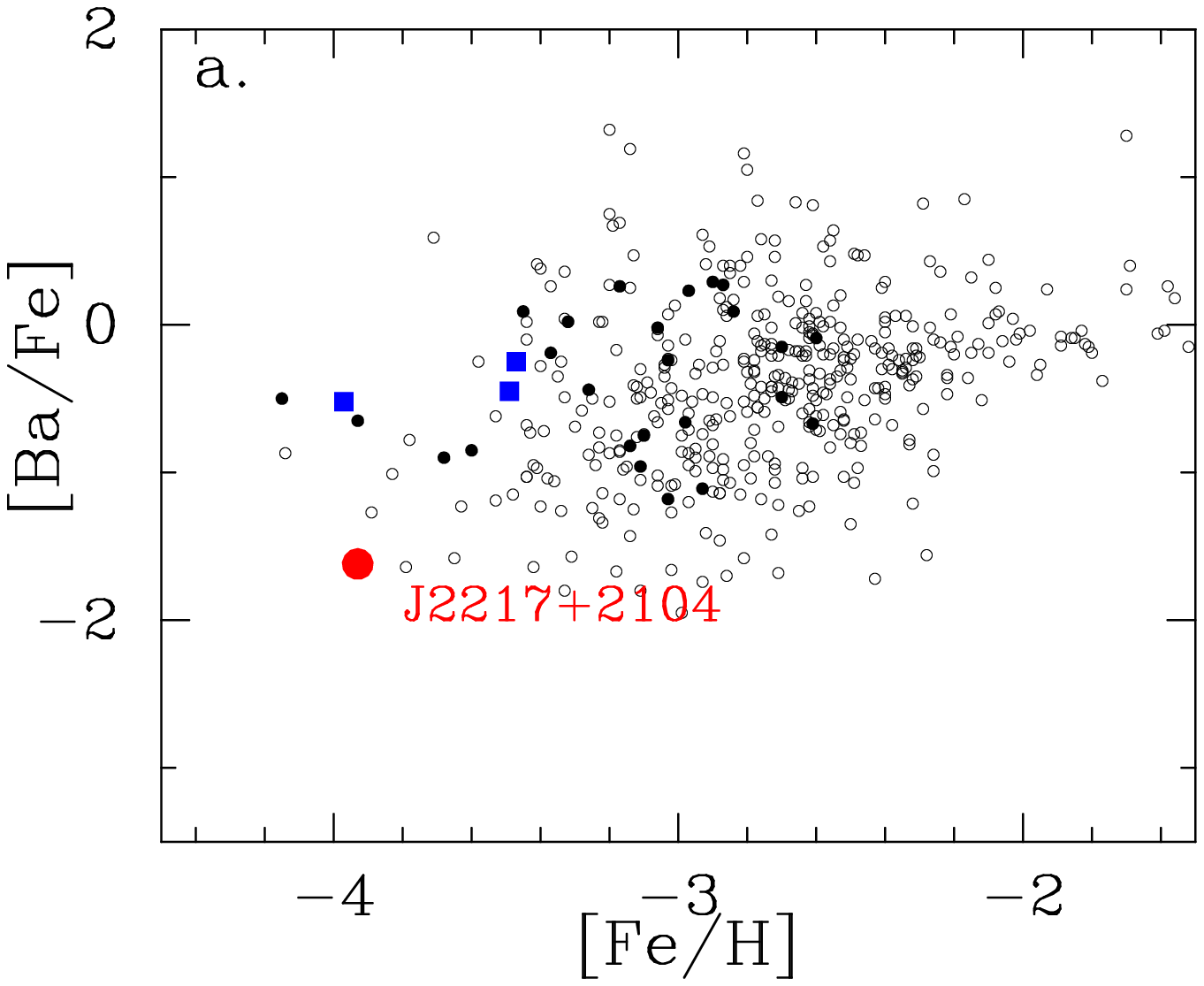}\\
  \includegraphics[width=8cm]{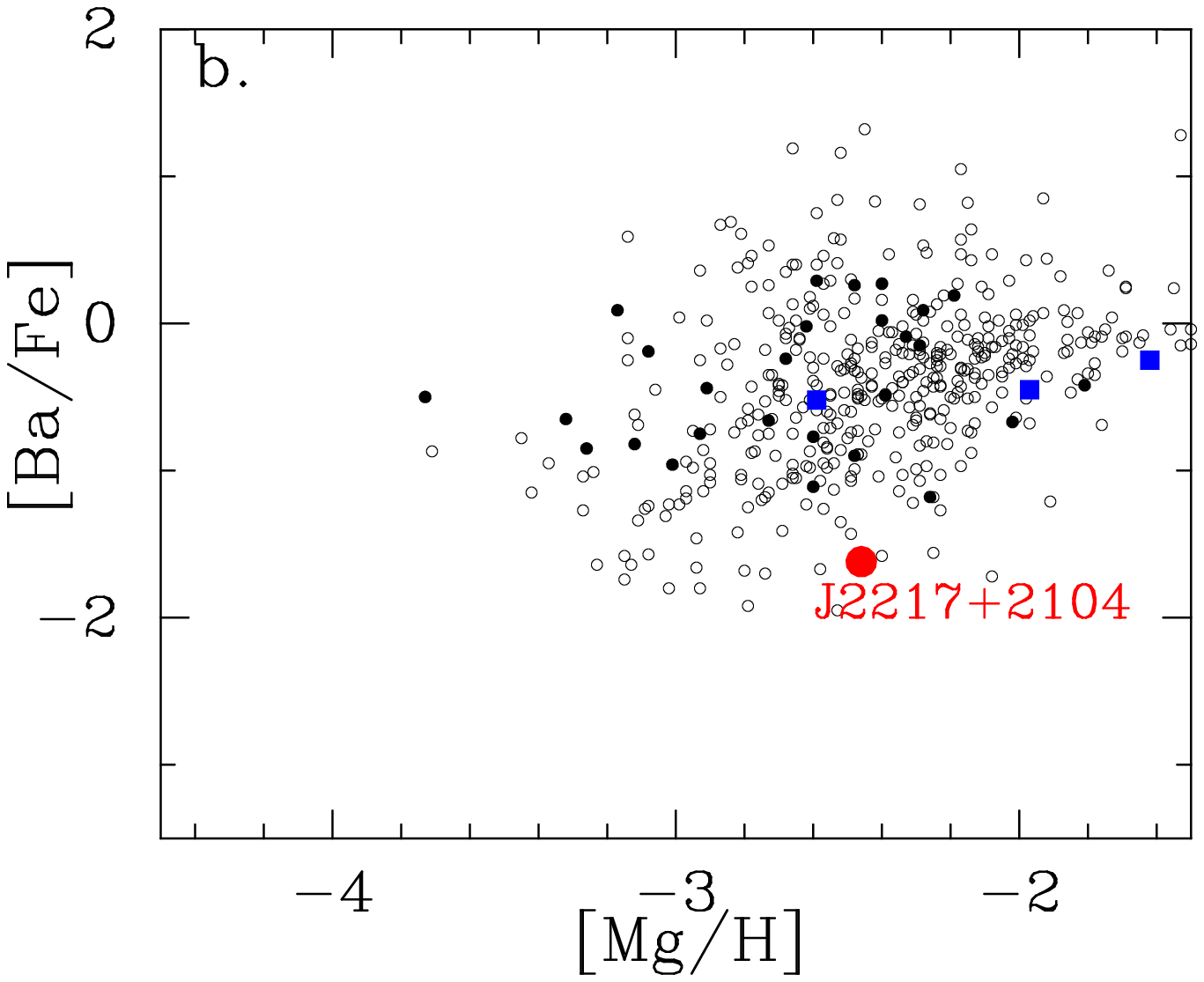}\\
  \includegraphics[width=8cm]{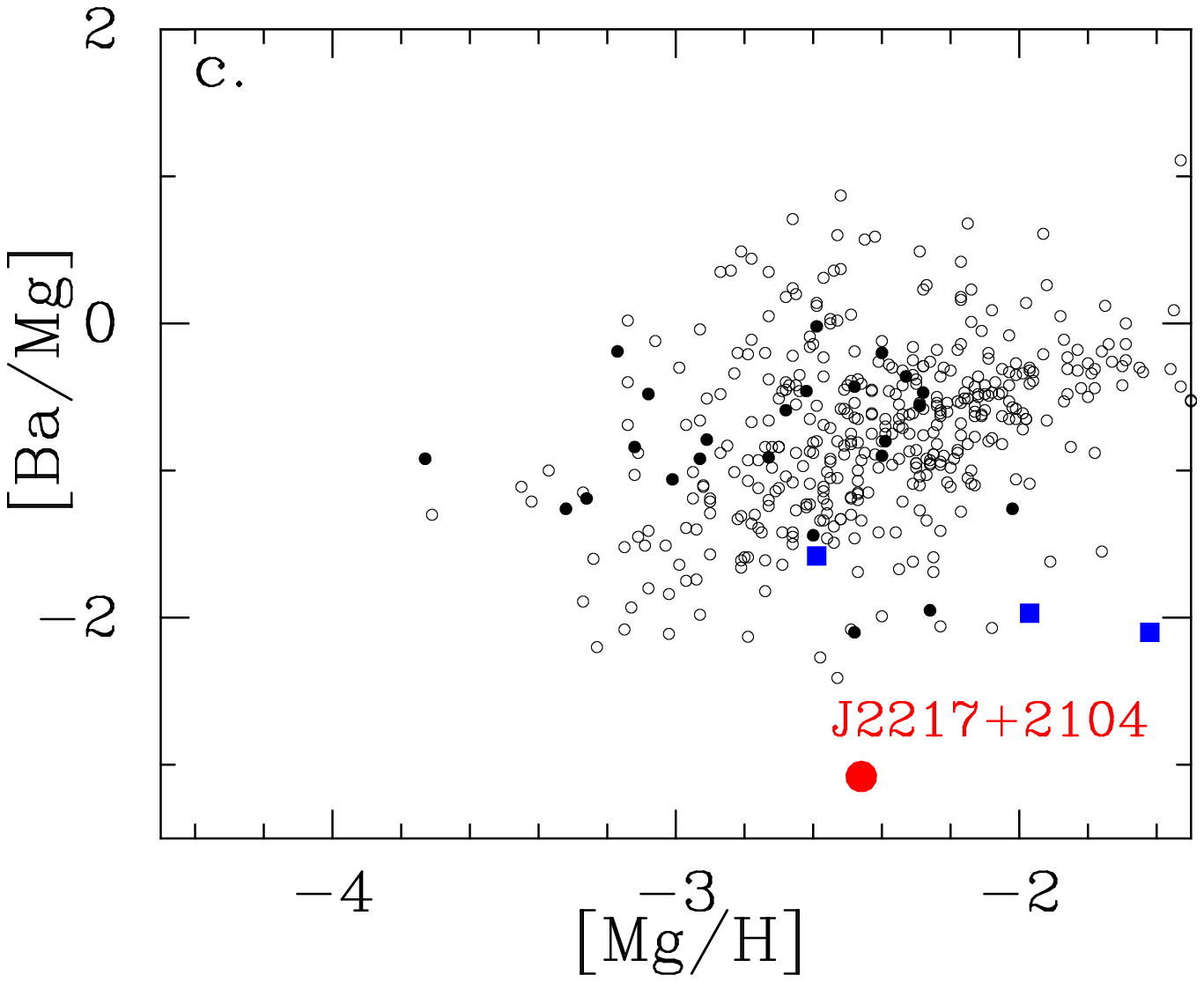}\\
 \end{center}
 \caption{[Ba/Fe] as a function of [Mg/H]. CEMP stars with excesses of Mg and Si are shown by the red filled circle (J~2217+2104) and blue filled squares. Others (small circles) are taken from literature as Figure 1: filled circles are CEMP-no stars and open circles are C normal stars or those for which C abundances are yet constrained.}\label{fig:bafe}
\end{figure}

Heavy neutron-capture elements are deficient in EMP stars with
[Fe/H]$<-3.5$ in general, in contrast to the objects with [Fe/H]$\sim
-3$ that show large star-to-star scatter in their abundance ratios
(e.g., [Eu/Fe]), including r-process-enhanced stars. In EMP stars with
low abundances of neutron-capture elements, measurable neutron-capture
elements are only Sr and Ba that have strong resonance lines in the
optical range. The [Sr/Fe] and [Ba/Fe] ratios in stars with
[Fe/H]$<-3.5$ are typically $-1$ (e.g., \cite{yong13}). Figure
  ~\ref{fig:bafe}a shows [Ba/Fe] as a function of [Fe/H] for
  metal-poor stars.  The [Ba/Fe] values of CEMP stars with excesses of
  Mg and Si previously reported are not as low as the typical [Ba/Fe]
  values of other EMP stars with similar Fe abundance. This was
discussed by \citet{aoki02} for CS~22949--037 and CS~29498--043.
J~2217+2104 has, however, very low abundances of Sr and Ba, as found
in C-normal stars with [Fe/H]$<-3.5$. This indicates that there exists
scatter in the abundance ratios of neutron-capture elements in such
CEMP stars.

For such CEMP stars, Fe is not necessarily a good indicator of
metallicity, because O, Mg, and Si are overabundant with respect to
Fe. Figure~\ref{fig:bafe}b shows [Ba/Fe] as a function of [Mg/H],
adopting Mg as an indicator of metallicity. CEMP-s stars are excluded
from this plot. Since the typical value of [Mg/Fe] in metal-poor stars
is +0.4, most stars shift to higher values in the horizontal axis
compared to the plot for [Fe/H]. The large scatter of [Ba/Fe] appears
at [Mg/H]$\sim -2.6$ in this diagram, corresponding to the scatter at
[Fe/H]$\sim -3$ in figure~\ref{fig:bafe}a. the usual plot. On the
other hand, CEMP stars with Mg excess shift by more than 1~dex. The
[Ba/Fe] values of the two particularly Mg-rich stars ([Mg/H]$>-2$) are
similar to the typical value in other stars, while the other two
stars, including J~2217+2104, are located within the scatter of
[Ba/Fe] found in other stars with [Mg/Fe]$\sim -2.6$.

This suggests that the neutron-capture elements of CEMP stars with
excesses of Mg and Si follow the distribution of neutron-capture
elements of other EMP stars showing no C-excess, and there is no
direct connection between the origin of excesses of C, N, O, Mg and Si
 and the event that has produced neutron-capture elements.

We note for completeness that, in the plot of [Ba/Mg] as a
  function of [Mg/H] shown in figure~\ref{fig:bafe}c, the CEMP stars
with excesses of Mg and Si are well separated from the other stars,
showing very low values of [Ba/Mg]. This is simply due to the large Mg
excess.

\subsection{Progenitors of CEMP stars with excesses of Mg and Si} 

The extremely low metallicity (low Fe abundance) of these objects
indicates that their elemental abundances would be predominantly
determined by the chemical yields of first generation stars. We find
49 objects in the range $-4.5<$[Fe/H]$<-3.5$ in figure~\ref{fig:mgfe}, among which
five are CEMP stars showing large Mg excesses, suggesting that the
frequency of such objects is in the order of 10\%. Hence, if the
chemical compositions of these stars are originated from similar
objects, e.g., some specific type of supernovae, they should not be
very rare in the early Galaxy. We note that the high fraction of
CEMP-no stars at very low metallicity might not only reflect the
frequency of their progenitors, but also depend on the efficiency of
low-mass star formation that is affected by chemical composition of the
gas cloud, as discussed by, e.g.,  \citet{norris13} and \citet{chiaki17}.

\begin{figure} 
\begin{center}
  \includegraphics[width=15cm]{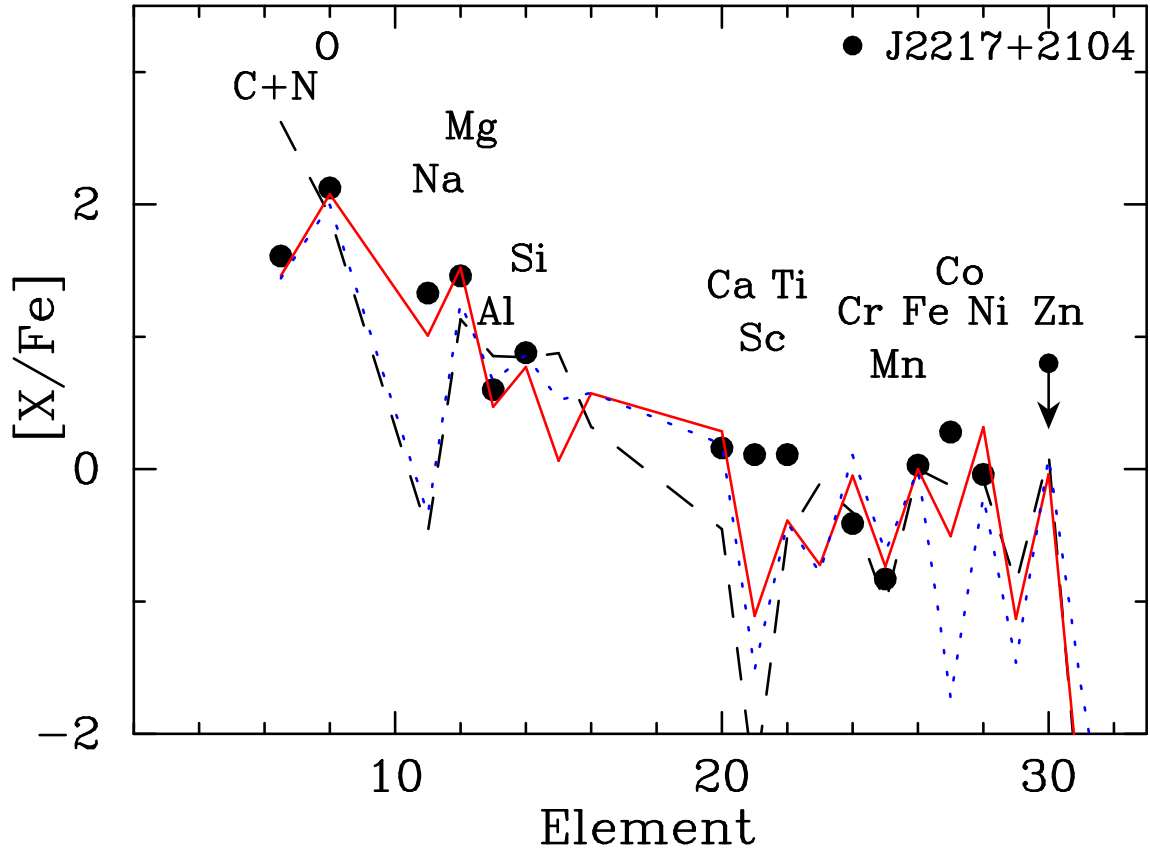}
 \end{center}
 \caption{The abundance pattern of J~2217+2104 and predictions by
   models of \citet{ishigaki18} for stellar masses of 13~M$_{\odot}$
   (black dashed line), 25~M$_{\odot}$ (red solid line) and
   40~M$_{\odot}$ (blue dotted line) with explosion energy of
   $1.0\times 10^{51}$~erg. The ratio of total abundances of C and N
   with respect to Fe ([(C+N)/Fe]) is shown. The model for
   $M=$25~M$_{\odot}$ well reproduces the abundance ratios from C+N to
   Ca.}\label{fig:model}
\end{figure}

A promising model proposed to explain the abundance patterns of these
objects is the so-called mixing-fallback model for supernova
explosions \citep{umeda03}. This model assumes significant mixing in
the inner region of a massive star during the explosion and unusually
large fall-back that results in little ejection of heavier elements
including Fe. Abundance patterns predicted by this model have been
compared with CEMP stars including CS~22949--037 and
CS~29498--043. Recently, \citet{ishigaki18} apply the models covering
wide ranges of masses and explosion energies to abundance patterns of
many EMP stars. They exhibit that the abundance patterns of CEMP stars
with excesses of Mg and Si are better explained by the models of
25~M$_{\odot}$ rather than those of more massive cases. This is in
particular constrained by the [Na/Mg] ratios. Figure~\ref{fig:model}
shows the abundance patterns predicted by models for three different
progenitor masses (13, 25 and 40 M$_{\odot}$) for explosion energy of
10$^{51}$ erg along with the abundance ratios of J~2217+2104. Here the
total abundances of C and N are presented taking account of possible
conversion of C to N by the CN-cycle after nucleosynthesis in a massive
progenitor. The abundance pattern of J~2217+2104 from C+N to Ca is well
reproduced by the model of 25~M$_{\odot}$, whereas the ratios of
(C+N)/Fe and Na/Fe are not explained by the other models.

The Co/Fe ratio is not well reproduced by any models, and is better
explained by models assuming large explosion energies (hypernova
cases). However, the [Na/Mg] values, as well as [(C+N)/O] values, of
the CEMP stars studied here are better explained by models with normal
explosion energy ($10^{51}$~erg; \cite{ishigaki18}). The explosion
energy is well constrained by the Zn abundance. The Zn abundance of
CS~22949--037 reported by \citet{depagne02} is [Zn/Fe]$=+0.7$,
suggesting a large explosion energy, whereas the upper limit derived
for CS~29498--043 by \citet{aoki04} is relatively low ([Zn/Fe]$<+0.5$)
and, hence, particularly high explosion energy is not required. The
upper-limit of the Zn abundance estimated for J~2217+2104,
[Zn/Fe]$<+0.7$, is still marginal, and further measurements will
provide useful constraints.  The high Sc and Ti abundances of
J~2217+2104 are not reproduced by any models, and the model values are
treated as lower limits in the fitting \citep{ishigaki18}. The
abundances of these two elements, as well as the high Co abundance,
could be explained by a jet-induced explosion \citep{tominaga14}.

The progenitor mass for these CEMP stars is also estimated to be about
20~M$_{\odot}$ by supernova models of
\citet{heger10}. \citet{placco16} compare the models with the
abundance pattern of CS~22949--037, which has a similar abundance pattern
to J~2217+2104, and concluded that the model for 21.5~M$_{\odot}$ of
\citet{heger10} provides the best fit. The strongest constraint is
also given by the abundance ratios of C(+N), O, Na and Mg. 

The initial mass function of first stars is one of the
most important issues to understand early structure formation. It has
been constrained by comparisons of models with observables like
metallicity distribution function of most metal-poor stars and the
fraction of CEMP stars (e.g., \cite{debennassuti17}). Estimating mass
of progenitor stars for individual metal-poor stars is another
approach to this goal. The mass of the progenitors estimated for CEMP
stars with excesses of Mg and Si is not particularly different from
that expected for C-normal EMP stars, This suggests that, although the
progenitors of such CEMP stars produced C-rich material due to some
properties of stars (e.g. rotation, binarity), they could represent
some important portion of initial mass function of first stars.

Significant mass-loss from first generation (metal-free) massive stars
with rapid rotation are also proposed to explain CEMP-no stars, in
particular the hyper metal-poor ([Fe/H]$<-5$) stars showing extremely large
excess of C and other light elements \citep{meynet06}. Remarkable
features predicted by this model are large excesses of odd elements,
i.e., N, Na, Al as a result of the CNO-cycle and efficient mixing
caused by rapid rotation. The large excesses of N and Na, as well as C
and Mg, found in the CEMP stars discussed here support the model. The
N excess might be, however, produced during the evolution of low-mass
stars we are currently observing, as discussed in
\S\ref{sec:cn}. Another difficulty is the excess of Si, which is not
expected in the material ejected from rotating massive stars.

Another model proposed as a possible origin of CEMP-no stars is mass
transfer from companion AGB stars in binary systems
\citep{suda04}. Large enhancements of Na and Mg are predicted by a
neutron-capture nucleosynthesis model of very metal-poor AGB stars
\citep{nishimura09}.
The large excess of Si found in the CEMP stars discussed here is,
however not explained by this model.
The [(C+N)/O] ratios observed for
the sample of this work are not as high as the C/O ratios expected
from AGB nucleosynthesis models. Hence, at lease according to the
current AGB nucleosynthesis models, mass transfer from AGB stars is
implausible to be the origin of CEMP-no stars with large excesses of
Mg and Si, whereas it could be a possible origin of some portion of
other CEMP-no stars.

\section{Summary and concluding remarks}

The elemental abundances of a newly discovered CEMP star, J~2217+2104,
are determined based on high resolution spectroscopy. This is a new
example of a class of CEMP stars having excesses of N, O, Na, Mg, Al
and Si. The abundance pattern of this class of objects from C to Ni is
very similar to each other, suggesting the existence of similar
progenitors in the early universe that would be a specific type of
supernova explosions. The abundance pattern of C, N, O, Na and Mg of
J~2217+2104 and other similar CEMP stars suggests that the progenitor
mass is about 25~M$_{\odot}$, which is not particularly different from
typical mass of progenitors expected for EMP stars in general. Hence,
the cause of the very different abundance pattern of these CEMP stars
from other EMP stars would be some properties of progenitors other
than their mass. The difference comes from the parameters of mixing
and fallback in the model \citep{ishigaki18}, which might be related to
the rotation, binarity, or other properties.

The kinematics information of Galactic stars is
becoming available by new astrometry data obtained with Gaia. The
sample size of this class is still too small to determine their
chemical and kinematic properties and their implication in the early
nucleosynthesis and star formation.  Ongoing and future surveys of
very metal-poor stars will be useful to search for EMP/UMP stars
including such carbon-enhanced objects.

\begin{ack}

This work is based on data collected at the Subaru Telescope, which is
operated by the National Astronomical Observatory of Japan.
Guoshoujing Telescope (the Large Sky Area Multi-Object Fiber
Spectroscopic Telescope, LAMOST) is a National Major Scientific
Project built by the Chinese Academy of Sciences.  Funding for the
project has been provided by the National Development and Reform
Commission.  LAMOST is operated and managed by the National
Astronomical Observatories, Chinese Academy of Sciences.
This work was supported by JSPS - CAS Joint Research Program.  WA and
TS were supported by JSPS KAKENHI Grant Numbers 16H02168, 16K05287 and
15HP7004.  HL was supported by NSFC grants No. 11573032, 11390371.
We thank Dr. M.Y. Fujimoto for useful comments on AGB nucleosynthesis.

\end{ack}

\clearpage

\begin{table}
\tbl{Abundance results}{
\begin{tabular}{llcccc} 
\hline\noalign{\vskip3pt}
Facility (resolving power) & Obs. date & HJD & Wavelengths & $S/N$ & $V_{\rm Helio}$ \\
\hline\noalign{\vskip3pt}
LAMOST ($R=2000$) & Oct. 23, 2013  & 2456588 & 3800-9000  & ... & $-116.7$ \\
Subaru ($R=45,000$) & Nov. 30, 2015 & 2457356.77 & 4030-6800  & 40 & $-115.79$ \\
Subaru ($R=60,000$) & Aug. 4, 2017 & 2457970.00 & 3500-5300  & 61 & $-116.13$  \\
Subaru ($R=60,000$) & Aug. 5, 2017 & 2457971.00 & 4030-6800  & 26 & $-115.29$  \\
Gaia DR2 &          &              &            &                & $-117.7$ \\
\hline\noalign{\vskip3pt} 
\end{tabular}}\label{tab:obs}%
\begin{tabnote}
The errors of LAMOST and Gaia results are 7~km~s$^{-1}$ and 3.3~km~s$^{-1}$, respectively.
\end{tabnote}
\end{table}

\begin{longtable}{lccccc} 
\caption{Atomic line data, equivalent widths and derived abundances}\label{tab:ew}
\hline
Species  & Wavelength ({\AA}) & L.E.P. (eV) & $\log gf$ & $W$(m{\AA}) & $\log \epsilon$ \\
\endhead
  \hline
\endfoot
  \hline
\endlastfoot
  \hline
Na I    &    5889.95 &   0.000 &    0.10 &  173.8 &   3.68 \\
Na I    &    5895.92 &   0.000 &   -0.20 &  151.4 &   3.60 \\
Mg I    &    4571.10 &   0.000 &   -5.69 &   80.5 &   5.19 \\
Mg I    &    4702.99 &   4.330 &   -0.44 &   60.7 &   4.99 \\
Mg I    &    5172.68 &   2.712 &   -0.45 &  195.4 &   5.23 \\
Mg I    &    5183.60 &   2.717 &   -0.24 &  216.1 &   5.23 \\
Mg I    &    5528.40 &   4.346 &   -0.50 &   62.6 &   5.04 \\
Al I    &    3961.52 &   0.014 &   -0.34 &  136.7 &   3.12 \\
Si I    &    4102.94 &   1.909 &   -3.14 &   48.5 &   4.46 \\
Ca I    &    4226.73 &   0.000 &    0.24 &  134.6 &   2.63 \\
Ca I    &    4454.78 &   1.898 &    0.26 &   21.3 &   2.51 \\
Sc II   &    4246.82$^{a}$ &   0.315 &    0.24 &  102.8 &  -0.45 \\
Sc II   &    4320.75$^{a}$ &   0.605 &   -0.25 &   35.7 &  -0.84 \\
Sc II   &    4415.56$^{a}$ &   0.595 &   -0.67 &   22.2 &  -0.73 \\
Ti I    &    4991.07 &   0.836 &    0.45 &   10.3 &   1.13 \\
Ti II   &    4417.72 &   1.165 &   -1.19 &   33.2 &   1.14 \\
Ti II   &    4443.80 &   1.080 &   -0.71 &   61.5 &   1.04 \\
Ti II   &    4450.48 &   1.084 &   -1.52 &   24.6 &   1.18 \\
Ti II   &    4464.45 &   1.161 &   -1.81 &   10.9 &   1.14 \\
Ti II   &    4468.49 &   1.131 &   -0.63 &   63.0 &   1.05 \\
Ti II   &    4501.27 &   1.116 &   -0.77 &   51.9 &   0.98 \\
Ti II   &    4533.97 &   1.237 &   -0.77 &   65.2 &   1.35 \\
Ti II   &    4563.77 &   1.221 &   -0.96 &   50.8 &   1.27 \\
Ti II   &    4571.97 &   1.572 &   -0.31 &   48.4 &   1.02 \\
Cr I    &    4289.72 &   0.000 &   -0.37 &   62.7 &   1.30 \\
Cr I    &    5206.04 &   0.941 &    0.02 &   23.2 &   1.27 \\
Cr I    &    5208.42 &   0.941 &    0.17 &   33.7 &   1.34 \\
Mn I    &    4033.06$^{a}$ &   0.000 &   -0.65 &   58.6 &   0.67 \\
Fe I    &    4063.59 &   1.558 &    0.06 &  104.9 &   3.29 \\
Fe I    &    4071.74 &   1.608 &   -0.01 &  104.2 &   3.40 \\
Fe I    &    4132.06 &   1.608 &   -0.68 &   85.9 &   3.58 \\
Fe I    &    4143.42 &   3.047 &   -0.20 &   15.7 &   3.37 \\
Fe I    &    4143.42 &   3.047 &   -0.20 &   15.7 &   3.51 \\
Fe I    &    4147.67 &   1.485 &   -2.10 &   22.2 &   3.53 \\
Fe I    &    4216.18 &   0.000 &   -3.36 &   70.1 &   3.84 \\
Fe I    &    4250.79 &   1.557 &   -0.71 &   72.6 &   3.21 \\
Fe I    &    4260.47 &   2.399 &    0.08 &   73.7 &   3.50 \\
Fe I    &    4271.76 &   1.485 &   -0.17 &  112.2 &   3.53 \\
Fe I    &    4282.40 &   2.176 &   -0.78 &   41.0 &   3.45 \\
Fe I    &    4337.05 &   1.557 &   -1.70 &   38.1 &   3.54 \\
Fe I    &    4375.93 &   0.000 &   -3.02 &   85.4 &   3.79 \\
Fe I    &    4383.54 &   1.485 &    0.21 &  118.5 &   3.27 \\
Fe I    &    4404.75 &   1.557 &   -0.15 &  104.6 &   3.36 \\
Fe I    &    4415.12 &   1.608 &   -0.62 &   89.1 &   3.51 \\
Fe I    &    4427.31 &   0.052 &   -2.92 &   98.8 &   4.07 \\
Fe I    &    4442.34 &   2.198 &   -1.25 &   30.4 &   3.74 \\
Fe I    &    4459.12 &   2.176 &   -1.28 &   24.2 &   3.60 \\
Fe I    &    4461.65 &   0.087 &   -3.21 &   77.8 &   3.91 \\
Fe I    &    4489.74 &   0.121 &   -3.97 &   29.0 &   3.82 \\
Fe I    &    4494.56 &   2.198 &   -1.14 &   24.5 &   3.49 \\
Fe I    &    4528.61 &   2.176 &   -0.82 &   40.7 &   3.47 \\
Fe I    &    4531.15 &   1.485 &   -2.15 &   24.8 &   3.62 \\
Fe I    &    4602.94 &   1.485 &   -2.21 &   23.4 &   3.64 \\
Fe I    &    4871.32 &   2.865 &   -0.36 &   19.4 &   3.38 \\
Fe I    &    4890.75 &   2.876 &   -0.39 &   16.2 &   3.33 \\
Fe I    &    4891.49 &   2.851 &   -0.11 &   22.9 &   3.21 \\
Fe I    &    4918.99 &   2.865 &   -0.34 &   22.3 &   3.43 \\
Fe I    &    4920.50 &   2.833 &    0.07 &   43.5 &   3.41 \\
Fe I    &    4957.60 &   2.808 &    0.23 &   54.4 &   3.40 \\
Fe I    &    4994.13 &   0.915 &   -2.96 &   21.9 &   3.61 \\
Fe I    &    5006.12 &   2.833 &   -0.61 &   17.8 &   3.54 \\
Fe I    &    5012.07 &   0.859 &   -2.64 &   45.6 &   3.68 \\
Fe I    &    5041.07 &   0.958 &   -3.09 &   20.2 &   3.75 \\
Fe I    &    5041.76 &   1.485 &   -2.20 &   25.2 &   3.64 \\
Fe I    &    5051.63 &   0.915 &   -2.80 &   36.3 &   3.74 \\
Fe I    &    5083.34 &   0.958 &   -2.96 &   25.8 &   3.74 \\
Fe I    &    5123.72 &   1.011 &   -3.07 &   16.9 &   3.69 \\
Fe I    &    5127.36 &   0.915 &   -3.31 &   12.4 &   3.65 \\
Fe I    &    5142.93 &   0.958 &   -3.08 &   16.3 &   3.62 \\
Fe I    &    5151.91 &   1.011 &   -3.32 &   12.7 &   3.80 \\
Fe I    &    5171.60 &   1.485 &   -1.79 &   43.7 &   3.57 \\
Fe I    &    5192.34 &   2.998 &   -0.42 &   16.3 &   3.50 \\
Fe I    &    5194.94 &   1.557 &   -2.09 &   25.6 &   3.62 \\
Fe I    &    5216.27 &   1.608 &   -2.15 &   16.9 &   3.52 \\
Fe I    &    5227.19 &   1.557 &   -1.23 &   73.4 &   3.58 \\
Fe I    &    5232.94 &   2.940 &   -0.06 &   30.9 &   3.42 \\
Fe I    &    5254.96 &   0.110 &   -4.76 &    8.7 &   3.91 \\
Fe I    &    5269.54 &   0.860 &   -1.32 &  121.3 &   3.76 \\
Fe I    &    5270.36 &   1.608 &   -1.51 &   61.4 &   3.72 \\
Fe I    &    5324.18 &   3.211 &   -0.10 &   12.7 &   3.31 \\
Fe I    &    5328.04 &   0.915 &   -1.47 &  110.5 &   3.72 \\
Fe I    &    5328.04 &   0.915 &   -1.47 &  110.5 &   3.52 \\
Fe I    &    5455.61 &   1.011 &   -2.10 &   71.2 &   3.69 \\
Fe I    &    5497.52 &   1.011 &   -2.85 &   24.7 &   3.64 \\
Fe I    &    5501.46 &   0.958 &   -3.05 &   17.9 &   3.61 \\
Fe I    &    5506.78 &   0.990 &   -2.80 &   31.3 &   3.70 \\
Fe I    &    5615.64 &   3.332 &    0.05 &   19.8 &   3.51 \\
Fe I    &    6230.72 &   2.559 &   -1.28 &    9.9 &   3.50 \\
Fe II   &    4522.63 &   2.844 &   -2.25 &   17.2 &   3.80 \\
Fe II   &    4923.93 &   2.891 &   -1.26 &   36.6 &   3.29 \\
Fe II   &    5018.45 &   2.891 &   -1.10 &   48.4 &   3.33 \\
Fe II   &    5234.63 &   3.221 &   -2.18 &   11.8 &   3.96 \\
Fe II   &    5316.62 &   3.153 &   -1.87 &   13.8 &   3.64 \\
Co I    &    4118.77$^{a}$ &   1.049 &   -0.48 &   41.2 &   1.56 \\
Co I    &    4121.31$^{a}$ &   0.922 &   -0.33 &   34.1 &   1.13 \\
Ni I    &    5476.90 &   1.826 &   -0.78 &   23.0 &   2.26 \\
Sr II   &    4077.71 &   0.000 &    0.15 &   53.6 &  -2.63 \\
Sr II   &    4215.52 &   0.000 &   -0.18 &   59.6 &  -2.21 \\
Ba II   &    4554.04 &   0.000 &    0.17 &   10.4 &  -2.21 \\
Ba II   &    4934.09 &   0.000 &   -0.16 &    9.9 &  -2.21 \\
\hline
\multicolumn{6}{l}{$^{a}$The effect of hyperfine splitting is included in the analysis.}
\end{longtable}

\begin{table}
\tbl{Abundance results obtained for J~2217+2104
\label{tab:abundance}}{%
\begin{tabular}{lcccccc} 
\hline\noalign{\vskip3pt}
 & $\log \epsilon$ & $n$ & $\log \epsilon{\odot}$ & [X/Fe] & $\sigma_{\rm [X/Fe]}$ \\
\hline\noalign{\vskip3pt}
C & 5.55        &   & 8.43 & 1.03 &  0.2 \\
N & 6.10        &   & 7.83 & 2.18 &  0.3 \\
O & 6.90        & 1 & 8.69 & 2.12 &  0.2 \\
Na & 3.64       & 2 & 6.24 & 1.33 & 0.18 \\
Mg & 5.14       & 5 & 7.60 & 1.46 & 0.08 \\
Al & 3.12       & 1 & 6.45 & 0.60 & 0.2  \\
Si & 4.46       & 1 & 7.51 & 0.88 & 0.18 \\
Ca & 2.57       & 2 & 6.34 & 0.16 & 0.15 \\
Sc II & $-0.67$ & 3 & 3.15 & 0.11 & 0.19 \\
Ti I & 1.13     & 1 & 4.95 & 0.11 & 0.19 \\
Ti II & 1.13    & 9 & 4.95 & 0.11 & 0.19 \\
Cr I & 1.30     & 3 & 5.64 & $-0.41$ & 0.11 \\
Mn & 0.67       & 1 & 5.43 & $-0.83$ & 0.19 \\
Fe I & 3.57    & 60 & 7.50 & $-3.93$ & 0.17 \\
Fe II & 3.60    & 5 & 7.50 & $-3.90$ & 0.24 \\
Co & 1.35       & 2 & 4.99 & 0.28 & 0.13 \\
Ni & 2.26       & 1 & 6.22 & $-0.04$ & 0.18 \\
Zn & $<1.5$     &   & 4.56 & $<0.8$ &  \\
Sr & -2.42      & 2 & 2.87 & $-1.37$ & 0.20 \\
Ba & -3.36      & 2 & 2.18 & $-1.62$ & 0.20 \\
\hline
$^{12}$C/$^{13}$C & $6^{+4}_{-2}$ & 3 & 89 & ... & ... \\
\hline\noalign{\vskip3pt} 
\end{tabular}
}
\end{table}

\begin{table}
\tbl{Abundance ratios of C, N, O, Na and Mg
\label{tab:ratios}}{%
\begin{tabular}{lccccccccccccc}
\hline\noalign{\vskip3pt}
  & $T_{\rm eff}$ & {$\log g$} & [Fe/H] & [C/Fe] & [N/Fe] & [O/Fe] & [Na/Fe] & [Mg/Fe] &  & [C/N] & [(C+N)/O] & [Na/Mg] & $^{12}$C/$^{13}$C\\
\hline\noalign{\vskip3pt}
J~2217+2104 & 4500 & 0.90 & -3.93 & 1.03 & 2.18 & 2.12 & 1.33 & 1.46 &  & -1.15 & -0.53 & -0.13 & 6 \\
CS~22949--037 & 4958 & 1.84 & -3.97 & 1.06 & 2.16 & 1.98 & 2.10 & 1.38 &  & -1.10 & -0.40 & 0.72 & 4  \\
CS~29498--043 & 4639 & 1.00 & -3.49 & 1.90 & 2.30 & 2.43 & 1.47 & 1.52 &  & -0.40 & -0.42 & -0.05 & 6 \\
HE~1012--1540 & 5745 & 3.45 & -3.47 & 2.22 & 1.25 & 2.25 & 1.93 & 1.85 &  & 0.97 & -0.12 & 0.08 & ... \\
HE~2139--5432 & 5416 & 3.04 & -4.02 & 2.59 & 2.08 & 3.15 & 2.15 & 1.61 &  & 0.51 & -0.63 & -0.06 & $>15$ \\
HE~1327--2326 & 6180 & 3.70 & -5.76 & 4.26 & 4.56 & 3.70 & 2.48 & 1.55 &  & -0.30 & 0.64 & 0.93 & $>5$ \\
\hline\noalign{\vskip3pt} 
\end{tabular}
}
\end{table}

\clearpage

\clearpage

\end{document}